\documentclass{article}

\usepackage{url}
\usepackage{graphicx}
\usepackage{natbib}
\usepackage{amsmath}

\RequirePackage{subcaption}

\begin{document}

\title{Parameter uncertainty in forecast recalibration}

\author{Stefan Siegert\footnote{Corresponding author address: s.siegert@exeter.ac.uk}, Philip G. Sansom and Robin Williams\\[2mm]
\multicolumn{1}{p{.8\textwidth}}{\centering\small\textit{Exeter Climate Systems, University of Exeter, Laver Building, North Park Road, Exeter, EX4 4QE, UK}}
}

\maketitle

\begin{abstract}
Ensemble forecasts of weather and climate are subject to systematic biases in the ensemble mean and variance, leading to inaccurate estimates of the forecast mean and variance.
To address these biases, ensemble forecasts are post-processed using statistical recalibration frameworks.
These frameworks often specify parametric probability distributions for the verifying observations.
A common choice is the Normal distribution with mean and variance specified by linear functions of the ensemble mean and variance.
The parameters of the recalibration framework are estimated from historical archives of forecasts and verifying observations.
Often there are relatively few forecasts and observations available for parameter estimation, and so the fitted parameters are also subject to uncertainty.
This artefact is usually ignored.                         
This study reviews analytic results that account for parameter uncertainty in the widely used Model Output Statistics recalibration framework.
The predictive bootstrap is used to approximate the parameter uncertainty by
resampling in more general frameworks such as Non-homogeneous Gaussian Regression.
Forecasts on daily, seasonal and annual time scales are used to demonstrate that accounting for parameter uncertainty in the recalibrated predictive distributions leads to probability forecasts that are more skilful and reliable than those in which parameter uncertainty is ignored.
The improvements are attributed to more reliable tail probabilities of the recalibrated forecast distributions.
\end{abstract}

\section{Introduction}
\label{sec:intro}

Raw forecasts produced by numerical atmosphere-ocean models are often not representative of the real world.
Modellers have long realised that there are systematic discrepancies between model simulations and the real world. 
\citet{glahn1972mos} suggested that a linear transformation should be applied to weather model forecasts to issue predictions of real world observables.
Similarly, \citet{leith1974theoretical} suggested that ``a final regression step is needed'' to get the ``best estimate of the true state''. 

The past 50 years has seen a shift in focus from forecasts that are deterministic in nature to probability forecasts that represent the forecaster's uncertainty of the future, while also providing a point forecast.
To this end, ensemble forecasts have become widely used in the fields of climate science and numerical weather prediction.
An ensemble forecast is simply a collection of forecasts that differ in one or more of the model physics, resolution, or initial conditions.

Statistical adjustment to better fit numerical model output to real world observations is also known as forecast recalibration.
A parametric statistical framework is specified that takes the raw numerical model forecast as an input, and outputs an estimate of the real world.
To avoid confusion due to overuse of the word ``model'', the term ``statistical framework'' is used in place of the more common ``statistical model''.
One of the simplest statistical forecast adjustments is the removal of a constant bias. 
The underlying assumption is that the observation is equal to the model output plus a constant offset. 
The offset is represented by an unknown parameter that must be estimated from training data.

A variety of more flexible recalibration frameworks have been proposed. 
In general, the choice of the recalibration framework depends on the forecast quantity, and on the assumptions about the forecast errors; for example, temperature forecasts require different recalibration frameworks than precipitation forecasts.
Two of the most well-known frameworks are Model Output Statistics \citep[MOS,][]{glahn1972mos} and Non-homogeneous Gaussian Regression \citep[NGR,][]{gneiting2005calibrated}. 
MOS is equivalent to Normal linear regression of the observations on the model output. 
NGR extends MOS by allowing the predictive uncertainty to depend on the ensemble spread. 
The more flexible the framework, the greater the number of parameters to be estimated.

It is common practice to recalibrate the forecast using the estimated parameters as the ``correct parameter values'', i.e., by treating them as known constants.
However, the parameters are estimated from a finite sample of training data and so are also subject to uncertainty. 
By na\"{\i}vely using the fitted parameters in issuing probability forecasts, the uncertainty in the parameter estimates is ignored.

The problem of accounting for parameter uncertainty when recalibrating climate and weather forecasts is not unknown.
\citet{glahn2009mos} state that the predictive distribution of forecasts recalibrated by linear regression should be a Student's t-distribution with inflated variance.
Related remarks can be found in \citet{mason2002comparison}, \citet{broecker2008from}, and \citet{unger2009ensemble}.
Forecast recalibration using dynamic linear models (Kalman filtering) also leads to forecasts that follow a t-distribution \citep{sohn2003the, pagowski2006application}.
Bayesian methods can also be used to account for parameter uncertainty in forecasts recalibrated by linear regression \citep{marty2014combining}, or by latent variable methods \citep{siegert2015bayesian}.

This study investigates the effect of accounting for parameter uncertainty on the reliability and skill of recalibrated probability forecasts.
Section~\ref{sec:methods} describes analytic methods to account for parameter uncertainty in MOS, and proposes a simple bootstrap method to account for parameter uncertainty in NGR.
Section~\ref{sec:results} applies the methods developed in Section~\ref{sec:methods} to three meteorological forecast data sets: an annual mean temperature forecast, a seasonal forecast of the North-Atlantic Oscillation, and a short-range 48-hour temperature forecast.
All three data sets demonstrate that accounting for parameter uncertainty improves both the skill and reliability of recalibrated forecasts.
Section~\ref{sec:summary} concludes with a summary and discussion.

\section{Methodology}
\label{sec:methods}

\subsection{Model Output Statistics (MOS)} 
\label{sec:MOS}

The original application of MOS was statistical downscaling, i.e., to produce forecasts of quantities that are not explicitly modelled numerically, such as surface winds or probability of precipitation \citep{glahn1972mos}.
However, MOS has been widely used for the statistical recalibration of both deterministic and ensemble forecasts \citep{kharin2003improved, tippett2005statistical, glahn2009mos}. 
As noted above, MOS is equivalent to Normal linear regression. 
The future (unknown) observation is represented by a linear function of the numerical model output, with Normally distributed forecast errors.
Let $y_t$ denote the observed conditions at time $t$, and let $m_t$ be the corresponding ensemble forecast mean.
Then the MOS recalibration framework is given by
\begin{equation}
  y_t = a + b m_t + c \varepsilon_t,
\end{equation}
where $\varepsilon_t$ is a standard Normally-distributed random variable, i.e., $\varepsilon_t\sim \mathcal{N}(0,1)$.
For simplicity, we focus on simple linear regression.
However, all the results presented in this paper extend easily to multiple linear regression, where the forecast mean depends on more than one input \citep[e.g.,][]{Glahn2009}.

The parameters $a$, $b$, and $c$ can be estimated by maximising the log-likelihood under the assumption that the errors $\varepsilon_t,t=1,2,\ldots,n$ are independent and follow a standard Normal distribution. 
Given a training set of $n$ ensemble forecast means $m_1, \dots, m_n$ and corresponding verifying observations $y_1, \dots, y_n$, the unbiased maximum-likelihood estimates of $a$, $b$, and $c^2$ are
\begin{subequations}
  \label{eqn:mosmle}
  \begin{align}
    \hat{a}   & = \bar{y} - \frac{s_{my}}{s^2_{m}}\bar{m},
    \label{eqn:mosmle-a} \\
    \hat{b}   & = \frac{s_{my}}{s^2_{m}},
    \label{eqn:mosmle-b} \\
    \hat{c}^2 & = \frac{1}{n-2} \sum_{t=1}^n 
                    \left( y_t - \hat{a} - \hat{b} m_t \right)^2,
    \label{eqn:mosmle-c}
  \end{align}
\end{subequations}
where $\bar{m}$ and $\bar{y}$ denote the overall sample means of the ensemble means $m_1, \dots, m_n$ and observations $y_1,\dots, y_n$ in the training sample, and $s_{my}$ and $s^2_m$ denote the sample covariance between ensemble means and observations and the sample variance of the ensemble means, respectively.
Note the division by $n-2$ in Eqn. \ref{eqn:mosmle-c}, to account for the fact that two mean parameters ($a$ and $b$) have been estimated \citep[][Chp. 1]{draper1998applied}.
The fitted parameters $\hat{a}$, $\hat{b}$ and $\hat{c}$ can then be used to recalibrate new forecasts for yet unknown observations.

\subsection{Non-homogeneous Gaussian Regression} 
\label{sec:NGR}

Recalibration using MOS explicitly assumes that the predictive variance is constant for all forecasts, i.e., equal to $c^2$.
In practice, the forecast uncertainty might be different on different occasions due to varying error growth rates, and more or less predictable weather regimes.
If the numerical model can reproduce this variability, then there might be useful information not only in the ensemble mean, but also in the ensemble variance. Recalibration using NGR aims to exploit systematic relationships between the ensemble variance and the variance of forecast errors \citep[NGR]{gneiting2005calibrated}. 
The NGR forecast mean is a linear function of the ensemble mean, $m_t$, identical to MOS.
But unlike MOS, the forecast variance at time $t$ is a linear function of the ensemble variance, $v_t$.
The NGR recalibration framework for the observation $y_t$ is thus
\begin{equation}
  y_t \sim \mathcal{N} \left( a + b m_t, c + d v_t \right).
  \label{eqn:ngr-model}
\end{equation}
Note that NGR recalibration with the parameter $d$ fixed at zero is equivalent to MOS recalibration.

\citet{gneiting2005calibrated} suggest parameter estimation for NGR by numerical minimisation of the Continuous Ranked Probability Score (CRPS).
However, \citet{williams2014comparison} found little advantage of minimum-CRPS estimation compared to maximum-likelihood estimation.
To maintain continuity with the MOS parameter estimators, the NGR recalibration parameters are estimated by maximising the log-likelihood function 
\begin{equation}
  \ell_{\text{NGR}} \propto - \sum_{t=1}^n \left[ 
    \log \left( c + dv_t \right) + 
      \frac{\left( y_t - a - b m_t \right)^2}{c + dv_t} 
  \right].
\end{equation}
Since closed form solutions are not available for the maximum-likelihood estimates of the NGR parameters, numerical optimisation is used.\footnote{Numerical optimisation is performed using the function {\tt optim} in the {\tt stats} package of the {\tt R} statistical programming environment \citep[version 3.2.0]{Rmanual}} 
To ensure that the forecast variance is positive, the variance scale parameter is represented by $d=\delta^2$ and optimised over $\delta$ \citep{gneiting2005calibrated}.

\subsection{Parameter uncertainty in MOS: Analytic results}
\label{sec:analytic}

Suppose the maximum-likelihood estimates $\hat{a}$, $\hat{b}$, and $\hat{c}$ from Equations~\ref{eqn:mosmle-a} -- \ref{eqn:mosmle-c} are known.
These parameter estimates can be used to recalibrate a new ensemble mean forecast $m^*$ for a yet unknown future observation $y^*$.
It is common practice to directly substitute the maximum-likelihood estimates into the MOS recalibration framework, so that the forecast distribution for $y^*$ is
\begin{equation}
  y^* \sim \mathcal{N} \left( \hat{a} + \hat{b} m^*, \hat{c}^2 \right).
  \label{eqn:predictive-normal}
\end{equation}
This recalibration framework has been used for probabilistic forecasting by, e.g., \citet{kharin2003improved, tippett2005statistical}. 

Results from linear regression \citep[][Chp. 1]{draper1998applied} show that the predictive distribution should also include parameter uncertainty. 
That is, the predictive distribution should not only include the estimated variance of the forecast errors $\hat{c}^2$, but should also account for the estimation uncertainty in the parameter estimates $\hat{a}$, $\hat{b}$ and $\hat{c}^2$.
The sampling variances of the parameter estimates $\hat{a}$ and $\hat{b}$ lead to an increase of the predictive variance of the Normal distribution.
Uncertainty due to estimation of the error variance $c^2$ leads to a transformation of the Normal distribution into a Student's t-distribution.

When parameter uncertainty is taken into account, the predictive distribution in linear regression becomes a t-distribution with inflated variance:
\begin{equation}
  y^* \sim t_{n-2} \left( \hat{a} + \hat{b} m^*, 
    \hat{c}^2 \left[ 1 + \frac{1}{n} + \frac{\left( m^* - \bar{m} \right)^2}
      {\sum_{t=1}^n \left( m_t - \bar{m} \right)^2} \right] \right).
  \label{eqn:predictive-t}
\end{equation}
The forecast variance is inflated by a term that depends on both the sample size $n$, and the distance of the ensemble forecast mean $m^*$ from the overall mean of the training forecasts, $\bar{m}$.
The function $t_\nu(\mu, \sigma^2)$ denotes the non-standardized Student's t-distribution with $\nu$ degrees of freedom, location $\mu$ and scale $\sigma$ (see appendix).
In the limit as $\nu \rightarrow \infty$, the t-distribution converges to a Normal distribution.
However, for small $\nu$, the tails of the t-distribution are heavier than those of the Normal distribution.
Therefore, the variances of the predictive distributions given by Equations~\ref{eqn:predictive-normal} and \ref{eqn:predictive-t} differ when the sample size $n$ is small, or when $\lvert m^* - \bar{m} \rvert$ is large.
Note that the forecast mean does not change.

\begin{figure}
  \centering
  \includegraphics[width=.8\textwidth]{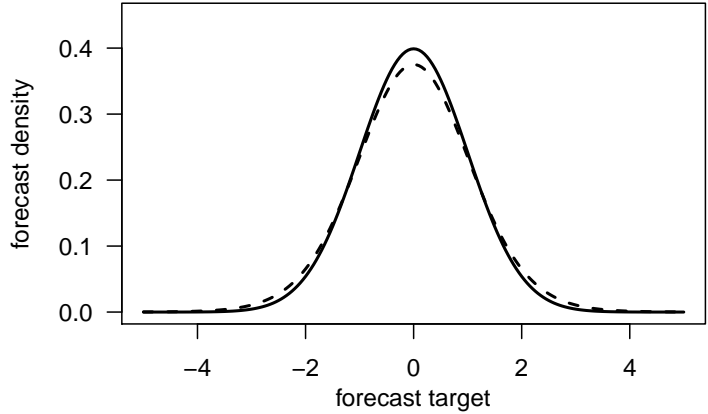}
  \caption{Illustration of difference between the standard Normal distribution $\mathcal{N}(0,1)$ (solid line) and the non-standardized t-distribution $t_{18}(0, 1.1)$ (dashed line).}
  \label{fig:normal-vs-t}
\end{figure}

The difference due to the adjustment for parameter uncertainty is illustrated in Figure~\ref{fig:normal-vs-t}. 
The standard Normal distribution $\mathcal{N}(0,1)$ is compared to the non-standardized t-distribution $t_{18}(0, 1.1)$.
The 18 degrees of freedom correspond to a sample size of $n=20$, and a 10\% inflated variance corresponds to a forecast $m^*$ that differs from the overall forecast mean of the training data by one standard deviation.
The example can thus be considered a typical case for what one might expect to see with a training sample of climate forecasts.
Figure~\ref{fig:normal-vs-t} shows that the difference due to the adjustment for parameter uncertainty is generally small, and so one should not expect the differences in forecast quality to be substantial.

\subsection{Parameter uncertainty in NGR: The predictive bootstrap}
\label{sec:bootstrap}

For NGR recalibration, \citet{gneiting2005calibrated} suggest substituting the parameter estimates $\hat{a},\hat{b},\hat{c}$ and $\hat{d}$, as well as the ensemble mean $m^*$ and variance $v^*$, directly into Equation~\ref{eqn:ngr-model}.
The forecast distribution for the future observation $y^*$ is then given by
\begin{equation}
  y^* \sim 
    \mathcal{N} \left( \hat{a} + \hat{b} m^*, \hat{c} + \hat{d} v^* \right).
  \label{eqn:ngr-pred}
\end{equation}
This approach has also been used to recalibrate probability forecasts in a number of other studies, e.g., \citet{hagedorn2008probabilistic, kann2009calibrating}.

However, in keeping with the discussion in Section~\ref{sec:analytic}, simply substituting the parameter estimates and issuing forecasts with a Normal distribution ignores parameter uncertainty.  
No analytic expression for the NGR forecast distribution that accounts for parameter uncertainty has been published, to date.
Therefore, a means of approximating the parameter uncertainty, and accounting for the parameter uncertainty in the predictive distribution is required.
The parameter uncertainty can be estimated non-parametrically by bootstrapping \citep{efron1982jackknife}.
Generating predictive distributions using bootstrap resampling is known as predictive bootstrapping, and was originally proposed by \citet{harris1989predictive}.
Given a historical archive of ensemble mean forecasts $m_t$, ensemble variances $v_t$, and corresponding observations $y_t$, for times $t=1,2,\ldots,n$,
the following bootstrap resampling protocol is proposed:
\begin{enumerate}
\item Generate a new training data set of size $n$ by randomly sampling $n$ times {\it with replacement} from the available pairs of historical forecasts and observations;
\item Compute the maximum likelihood NGR parameter estimates using this new training data set; and
\item Repeat steps 1. and 2. $K$ times.
\end{enumerate}
The $k$th resampling leads to the bootstrap parameter estimates $\tilde{a}_k$, $\tilde{b}_k$, $\tilde{c}_k$, and $\tilde{d}_k$. 
The collection of $K$ bootstrap parameter estimates approximates the parameter uncertainty distribution.

The objective is to generate a recalibrated forecast for the unknown value $y^*$ of the observation, using the ensemble mean forecast $m^*$ and ensemble variance $v^*$.
Each set of bootstrapped parameter estimates $(\tilde{a}_k,\tilde{b}_k,\tilde{c}_k,\tilde{d}_k),k=1,2,\ldots,K$ leads to a Normally-distributed forecast given by Equation~\ref{eqn:ngr-pred}.
The $K$ bootstrap samples are combined into a single predictive distribution by calculating the equally weighted average over the individual Normal distributions, thereby producing a Normal mixture distribution (see appendix).
The bootstrap forecast distribution itself is therefore not a Normal distribution.

\begin{figure}
  \begin{center}
  \includegraphics[width=.8\textwidth]{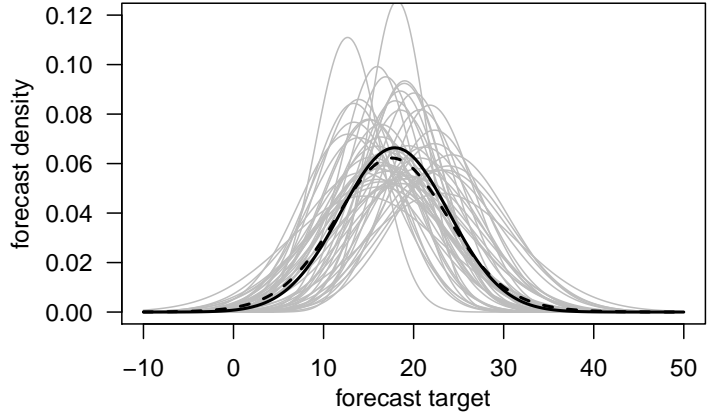}
  \end{center}
  \caption{Illustration of the predictive bootstrap to account for parameter uncertainty in NGR using a training data set of $n=19$. The Normal forecast distribution using maximum likelihood estimates of the NGR parameters (solid black line), 50 Normal distributions with maximum likelihood parameter estimates obtained by resampling the training data set (solid gray lines), and the distribution obtained by the averaging the 50 bootstrap distributions (dashed black line).} 
  \label{fig:normmix-illustration}
\end{figure}

The predictive bootstrap is illustrated in Figure~\ref{fig:normmix-illustration}.
The effect of averaging over the bootstrapped Normal distribution is similar to the adjustment for parameter uncertainty in MOS: The variance of the distribution is increased, and the tails are made heavier.
Unlike MOS, the mode of the predictive distribution can be different after accounting for parameter uncertainty.
By exploring different values for the estimated slope parameter $\hat{b}$, bootstrapping also inflates the forecast variance when the forecast mean $m^*$ is far from the mean of the training data $\bar{m}$.

The assumptions underlying the predictive bootstrap are quite different from those of the analytic method in Section~\ref{sec:analytic}. 
Both approaches assume that the training data represent independent and identically distributed samples (conditional on the parameters) from some unknown distribution.
However, the bootstrap approximates the distribution of the parameters, without making prior assumptions about the form of the distribution, e.g., Normal, Student t, etc..
Therefore, bootstrapping is useful for frameworks such as NGR, where parametric inference is difficult or impossible.

\subsection{Forecast verification: CRPS, Ignorance, PIT histogram}
\label{sec:verification}

This section outlines the forecast verification measures used in this paper to assess the quality and improvements of probabilistic forecasts. 
Only the general forms of the verification measures are provided here. 
Equations for specific distributions can be found in the appendix.

The Ignorance score is a verification score to evaluate probability density forecasts \citep{roulston2002evaluating}.
If the forecast probability density function (pdf) is $f(x)$, and the verifying observation is $y$, then the Ignorance score is given by
\begin{equation}
  ign(f, y) = - \log_2 f(y),
\end{equation}
i.e., the negative logarithm of the forecast density evaluated at the observation.
The Ignorance score is a local score, i.e. it only depends on the value of the forecast assigned to the verifying observation.
If the basis 2 is used, Ignorance differences are measured in bits.
An Ignorance difference of $\Delta>0$ bits between forecast A and forecast B implies that forecast B has assigned $2^\Delta$ times more probability density to the verifying observation than forecast A.
Forecast B, having lower Ignorance score than forecast A, can thus be considered to be the ``better'' forecast.
Time-averaged Ignorance differences are used as summary measures of relative forecast performance.

The Continuous Ranked Probability Score (CRPS) is designed to evaluate a cumulative forecast distribution (cdf) $F(x)$ for a scalar observation $y$ \citep{matheson1976scoring, hersbach2000decomposition}.
The general form of the CRPS is
\begin{equation}
  crps(F, y) = \int_{-\infty}^{+\infty} \left[ F(x) - H(x - y) \right]^2 dx,
\end{equation}
where $H(x)$ is the Heaviside step-function, i.e., $H(x) = 0$ for $x \le 0$ and $H(x)=1$ otherwise.
Unlike the Ignorance score, the CRPS is a non-local score. 
The CRPS depends not only on the value assigned to the observation, but also on how much forecast probability is concentrated near the observation, i.e., the CRPS is sensitive to the distance of the bulk of the forecast distribution from the observation.
For a deterministic forecast (i.e. where $F(x)$ is itself a step function), the CRPS is equal to the absolute difference between forecast and observation, and the CRPS vanishes for a perfect deterministic forecast where $F(x) = H(x-y)$.
Therefore, the CRPS can be interpreted as a distance measure between forecast and observation, and a lower CRPS value can be taken as an indication of a ``better'' forecast.
Note that the notion of ``better'' depends on the score, e.g., forecast A can perform ``better'' than forecast B in terms of the Ignorance score, but ``worse'' in terms of the CRPS. 

The CRPS is measured on the scale of the forecast target $x$.
The Continuous Ranked Probability Skill Score (CRPSS) provides a dimensionless measure of the difference in forecast performance.
If the time-averaged CRPS of forecasts A and B are $CRPS_A$ and $CRPS_B$, the relative improvement of forecast $B$ over forecast $A$ is given by
\begin{equation}
  CRPSS = \frac{CRPS_A - CRPS_B}{CRPS_A}.
\end{equation}
A positive (negative) CRPSS indicates an improvement (deterioration) of forecast B compared to forecast A.
CRPSS close to zero indicates that the forecasts are equally good. The CRPSS is
bounded above at unity for a perfect forecast B, but has no lower bound.

Probabilistic forecasts should be issued such that the verifying observations behave like random draws from the forecast distributions. 
Such forecasts are referred to as being reliable or well-calibrated \citep{gneiting2007probabilistic}.
For reliable forecasts, the observations fall on average equally often between the 0 and 5 percentiles, between the 5 and 10 percentiles, etc., of the forecast distribution.
Therefore, counting how often each percentile interval is occupied by the observation provides a simple test of forecast reliability.
In practice, the probability integral transform (PIT) of each forecast pdf $f_t(x)$ is calculated, which is the mass of forecast probability below its verifying observation:
\begin{equation}
  pit_t := PIT(f_t, y_t) = \int_{-\infty}^{y_t} f_t(x) dx.
\end{equation}
For example, if the observation $y_t$ falls between the 5th and 10th percentiles of the forecast density $f_t(x)$, then the value of $pit_t$ will be between $0.05$ and $0.10$.
Since each percentile interval should be equally likely on average for a reliable forecast, reliability can be checked by observing the shape of the PIT histogram. 
Reliable forecasts have a flat PIT histogram, and a non-flat PIT histogram indicates unreliable forecasts.
In general, $\cup$-shaped histograms indicate underdispersed forecast distributions, i.e., overconfident forecasts.
Conversely, $\cap$-shaped histograms suggest overdispersed forecast distributions (underconfident forecasts), while sloping histograms are indicative of a systematic bias of the forecast mean \citep{hamill2001interpretation}.

\section{Results}
\label{sec:results}

\subsection{CanCM4 gridded annual temperature forecasts}
\label{sec:cancm4}

The effect of accounting for parameter uncertainty was evaluated in forecasts of near-surface (2m) temperature generated by the fourth version of the Canadian coupled ocean-atmosphere general circulation model \citep[CanCM4;][]{cancm4}.
Forecast ensembles of 10 initial condition members were initialized on 1 January every year between 1960 and 2010. The forecast target was the mean temperature averaged over the first 12 months after initialization.
Verifying observations were taken from the HadCRUT3v data set \citep{brohan2006hadcrut3}.
Only grid boxes with a complete observational record were included in the analysis.
The conclusions were not found to be sensitive to the choice of verification data; comparable results were obtained when verifying against the ERA-Interim reanalysis \citep{dee2011era} and all grid boxes.
MOS-recalibrated probability forecasts were computed for each year in the period 1991--2010.
It was found that the ensemble variance was not a skilful predictor of forecast errors, and so there was no advantage to using NGR recalibration.  
Forecasts were recalibrated using only information available up to the time of the forecast and evaluated out-of-sample. 
The recalibration parameters for each forecast were estimated using the previous 25 years of forecasts and observations, after linearly detrending both data sets. 

\begin{figure*}[!ht]
  \centering
  \includegraphics[width=\textwidth]{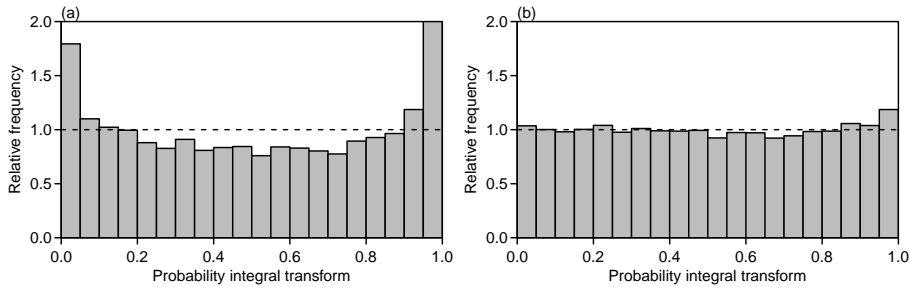}
  {\phantomsubcaption \label{fig:cancm4-pit-normal}}
  {\phantomsubcaption \label{fig:cancm4-pit-t}}
  \caption{PIT histograms of recalibrated CanCM4 forecasts
           (\subref{fig:cancm4-pit-normal})  
           before, and 
           (\subref{fig:cancm4-pit-t}) 
           after accounting for parameter uncertainty.
           Before accounting for parameter uncertainty, the observations fall into the tails of the forecast distribution too often.}
  \label{fig:cancm4-pit}
\end{figure*} 

The PIT histogram of the recalibrated forecasts without parameter uncertainty (Equation~\ref{eqn:predictive-normal}) appear $\cup$-shaped, indicating underdispersed probability forecasts (Figure~\ref{fig:cancm4-pit-normal}).
As a result of the underdispersion, the 90\% prediction intervals cover the verifying observations only 81\% of the time. 
Therefore, the forecasts without parameter uncertainty are not well calibrated.
In contrast, the PIT histogram of the recalibrated forecasts after accounting for parameter uncertainty (Equation~\ref{eqn:predictive-t}) is almost completely flat (Figure~\ref{fig:cancm4-pit-t}).
Each 5\% interval of the predictive distributions covers the verifying observation roughly 5\% of the time, and the $90\%$ prediction intervals cover the verifying observations $89\%$ of the time.
The recalibrated forecasts including parameter uncertainty appear to be well calibrated.

\begin{figure*}[!ht]
  \centering
  \includegraphics[width=\textwidth]{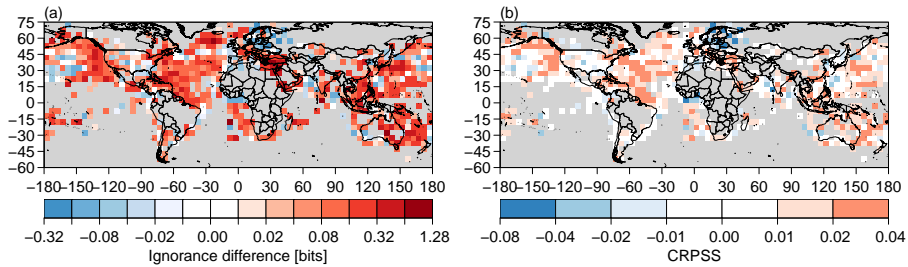}
  {\phantomsubcaption \label{fig:cancm4-igndiff}}
  {\phantomsubcaption \label{fig:cancm4-crpss}}
  \caption{Improvement in overall skill of the recalibrated CanCM4 forecasts after accounting for parameter uncertainty, measured by 
           (\subref{fig:cancm4-igndiff})  
           the difference in the Ignorance score, and 
           (\subref{fig:cancm4-crpss}) 
           the CRPSS.
           Positive score differences / skill scores indicate improved average scores after accounting for parameter uncertainty.}
  \label{fig:cancm4-skill}
\end{figure*} 

The time-averaged difference in the Ignorance score between the recalibrated forecasts before and after accounting for parameter uncertainty is positive at most grid boxes (Figure~\ref{fig:cancm4-skill}\subref{fig:cancm4-igndiff}).
Positive Ignorance differences indicate that the forecasts that account for parameter uncertainty assign more probability density to the observations than those that do not, i.e., are more skilful. 

The CRPSS is also positive in the majority of grid boxes, supporting the conclusion that the forecasts that account for parameter uncertainty are on average more skilful (Figure~\ref{fig:cancm4-skill}\subref{fig:cancm4-crpss}).

\subsection{Met Office seasonal NAO forecasts}
\label{sec:nao}

\begin{figure}
  \centering
  \includegraphics[width=.8\textwidth]{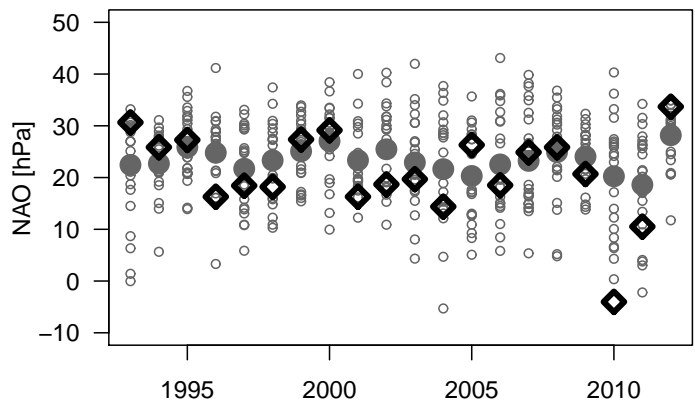}
  \caption{GloSea5 NAO ensemble forecasts (small circles), ensemble mean forecasts (large filled circles) and verifying observations (black diamonds) plotted over the verification year.}
  \label{fig:glosea5}
\end{figure}

This section presents a case study on seasonal climate forecasts using a more complicated recalibration framework.
The data consist of 20~years of historical seasonal ensemble forecasts of the North Atlantic Oscillation (Figure~\ref{fig:glosea5}). 
The ensembles were produced by the UK Met Office Global Seasonal prediction system GloSea5 \citep{maclachlan2014global}.
The forecasts were initialised using a lagged initialisation around 1 November each year between 1992 and 2011.
The forecast target is the average North Atlantic Oscillation (NAO) index between December and February (DJF), measured as a pressure difference between stations situated in the Azores and Iceland \citep{scaife2014skillful}.

\begin{figure}
  \centering
  \includegraphics[width=.8\textwidth]{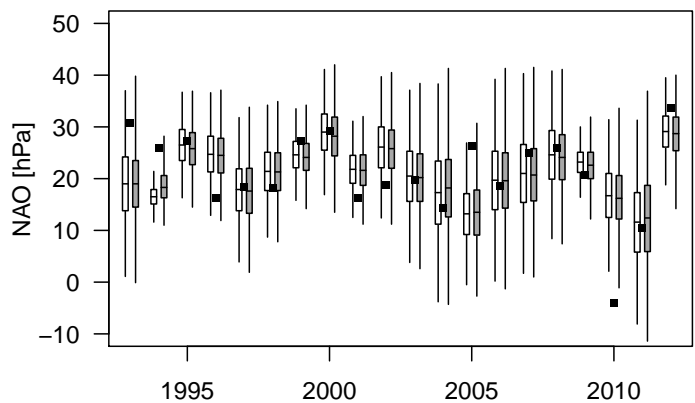}
  \caption{NAO observations (black markers) and their NGR predictive distributions using the ensemble data shown in Figure~\ref{fig:glosea5}. Distributions without parameter uncertainty (white boxes) and with parameter uncertainty (gray boxes) are depicted by box-and-whiskers plots, where boxes indicate the inter-quartile range and median, and whiskers extend from the 1 to the 99 percentile.}
  \label{fig:glosea5-pred}
\end{figure}

The sample correlation coefficient between the ensemble means and the observations is 0.62. 
The correlation between the ensemble standard deviation and absolute error of the ensemble mean is also high at 0.45 (or 0.3 when the very influential year 2010 is excluded).
Previous studies found that the skill of these forecasts can be improved by linear transformations of both the ensemble mean and spread \citep{eade2014}.
Therefore, NGR recalibration was used to allow for linear adjustment not only of the predictive mean, but also the predictive variance. Due to the small sample size, forecasts were evaluated by leave-one-out cross-validation, i.e., each forecast was recalibrated using the other 19 as the training set. Due to the long time scale under consideration, each forecast occasion can reasonably be assumed to be independent, and so the leave-one-out approach is justifiable.

Figure~\ref{fig:normmix-illustration} illustrates the predictive distributions for the year 1997, issued as a Normal distribution with the maximum likelihood NGR parameter estimates, and issued as a mixture of Normal distributions based on 500 bootstrap replicates.
The Normal distributions generated from the individual bootstrap resamples vary considerably, indicating large parameter uncertainty.
The variance of the averaged bootstrap distribution is larger, and the tails are heavier than for the Normal distribution that uses only the maximum-likelihood parameters.
Figure~\ref{fig:glosea5-pred} shows the recalibrated predictive distributions for the years 1993-2012, with and without parameter uncertainty, and their verifying observations.
The forecast distributions with parameter uncertainty have slightly heavier tails, and their medians are on average slightly closer to the climatological mean.

\begin{figure}
  \centering
  \includegraphics[width=.8\textwidth]{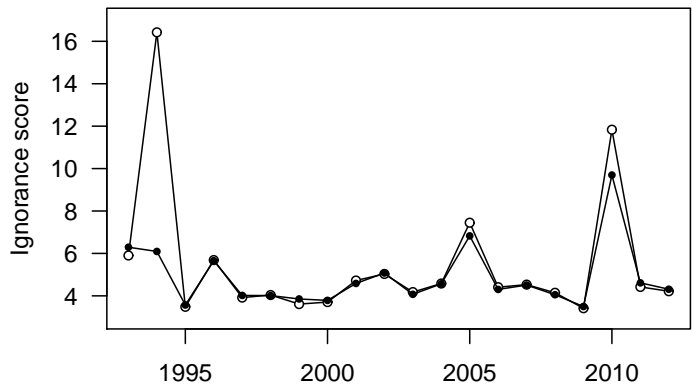}
  \caption{Ignorance scores of the forecast distributions shown in Figure~\ref{fig:glosea5-pred}, without accounting for parameter uncertainty (open circles) and after accounting for parameter uncertainty (full circles).}
  \label{fig:glosea5-ign}
\end{figure}

Figure~\ref{fig:glosea5-ign} shows the Ignorance scores of the individual forecasts.
Whenever the observation falls close to the bulk of the forecast distributions, the Ignorance scores for the two distributions are very similar.
There are three cases where the observation falls well into the tail of the forecast distributions (1994, 2005 and 2010). 
For these tail events, the Ignorance score improves when parameter uncertainty is taken into account.
The predictive bootstrap leads to heavier tails, and thus to higher probabilities being assigned to ``unexpected events''.

The results of the case study presented in this Section suggest that the predictive bootstrap increases the forecast skill of recalibrated GloSea5 NAO forecasts.
However, only 20 data points are considered, which does not allow for a robust statistical analysis of the forecast skill.
A larger dataset of numerical weather predictions is analysed in the next Section.

\subsection{NCEP short range temperature forecasts} 
\label{sec:NCEP}

Daily forecasts of near-surface (2m) temperature with a 48 hour lead time were also analysed.  
The forecasts were taken from version 2 of the reforecast project, hosted by the National Oceanic and Atmospheric Administration, USA \citep{hamill2013noaa}.  
The ensemble forecasts are approximately equivalent to those issued by the operational global ensemble forecasting system of the national centre for environmental prediction (NCEP). 
Ten member initial condition forecasts were issued at 00 UTC each day for a grid point close to New York City, USA (40N, 74W).
The forecasts covered the period 26 May 1990 - 15 September 2014, giving a total of $n=8,879$ forecasts and verifying observations. 
The analyses (i.e., the control forecast at 0 lead time) were used as verifying observations.

Preliminary investigations showed that NGR recalibration yielded more skilful forecasts than simple MOS recalibration.
Therefore, the NGR recalibrated forecasts were used throughout.
Recalibration parameters were estimated separately for all $n$ forecasts, using data from a rolling training window of pre-specified size $w$.
That is, for each forecast, the $w$ previous pairs of forecasts and observations were used as training data, such that all forecasts were evaluated out-of-sample.
The rolling training window allows the recalibration scheme to adapt to non-stationarities in the hindcast data, such as the updating of the data assimilation scheme in 2011, or the dependency of the forecast bias on the time of the year \citep{hamill2013noaa}.
Parameter uncertainty was accounted for using the bootstrap approach described in Section~\ref{sec:bootstrap}. 50 bootstrap replicates were used for each forecast. Increasing the number of bootstrap replicates was found to provide only very small improvements in forecast skill, at considerable computational expense due to the large sample of forecasts to be evaluated.

\begin{figure*}[!ht]
  \centering
  \includegraphics[width=\textwidth]{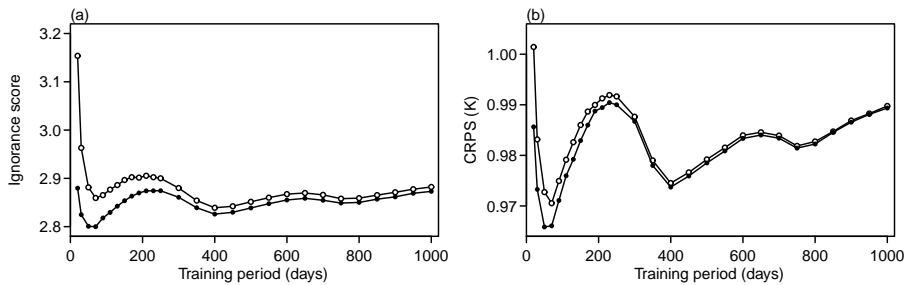}
  {\phantomsubcaption \label{fig:ign_trainingsize_ncep}}
  {\phantomsubcaption \label{fig:crps_trainingsize_ncep}}
  \caption{(\subref{fig:ign_trainingsize_ncep}) Ignorance scores, and 
           (\subref{fig:crps_trainingsize_ncep}) CRPS 
           as a function of training sample size for the recalibrated NCEP forecasts before (open circles) and after (filled circles) accounting for parameter uncertainty.}
  \label{fig:trainingsize-ncep}
\end{figure*} 

Figure~\ref{fig:trainingsize-ncep}\subref{fig:ign_trainingsize_ncep} shows the Ignorance scores of the NGR-recalibrated forecasts as a function of the size of the rolling training window, before and after accounting for parameter uncertainty. 
The improvements in forecast skill produced by accounting for parameter uncertainty are evident for both small and large training samples. 
As expected, the scores converge for very large training datasets.
It is encouraging that the predictive bootstrap yields improved forecasts even in relatively data-rich settings with hundreds of historical forecasts and observations, where one might expect the effect of parameter uncertainty to be negligible.
The CRPS values shown in Figure~\ref{fig:trainingsize-ncep}\subref{fig:crps_trainingsize_ncep} are qualitatively similar to those of the Ignorance scores.
The optimum training period after accounting for parameter uncertainty is 50 days for both scores.
Before accounting for parameter uncertainty, the optimal training period for CRPS is similar (60 days).
However, the optimal training period for the Ignorance score is 400 days without accounting for parameter uncertainty.  
For CRPS, the effect of the training length is large compared to the effect of parameter uncertainty, while the effects are of comparable magnitude for the Ignorance score. 

\begin{figure*}[!ht]
  \centering
  \includegraphics[width=\textwidth]{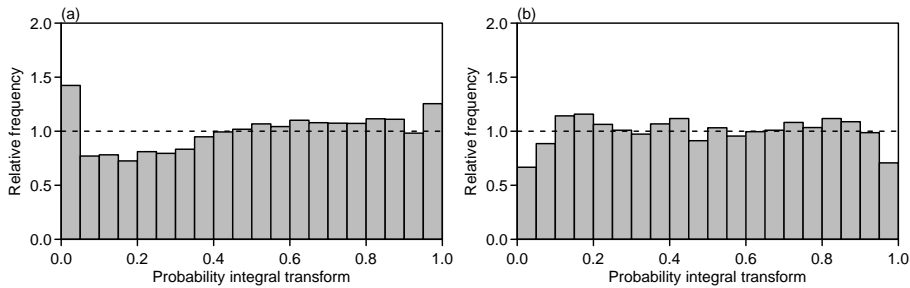}
  {\phantomsubcaption \label{fig:ncep-pit-mle}}
  {\phantomsubcaption \label{fig:ncep-pit-boot}}
  \caption{PIT histograms of recalibrated NCEP forecasts
           (\subref{fig:ncep-pit-mle})  before, and 
           (\subref{fig:ncep-pit-boot}) after 
           accounting for parameter uncertainty.
           Before accounting for parameter uncertainty, the forecasts show evidence of underdispersion and bias.
           After accounting for parameter uncertainty, the forecasts appear slightly overdispersed.}
  \label{fig:ncep-pit}
\end{figure*} 

Figures~\ref{fig:ncep-pit}\subref{fig:ncep-pit-mle} and \subref{fig:ncep-pit-boot} show PIT histograms after recalibration based on a rolling training period of 50 days, before and after accounting for parameter uncertainty.
The overpopulation of the outer bins of the PIT histograms when parameter uncertainty was not accounted for is indicative of forecast distributions whose tails are too light.
Accounting for parameter uncertainty by bootstrap resampling results in PIT histograms that are closer to being uniform.
Figure~\ref{fig:ncep-pit}\subref{fig:ncep-pit-boot} suggests that the resampling method has slightly overcompensated for the light tails, thus leading to overdispersive forecasts.
In both cases, the PIT histograms of the bootstrap forecasts suggest remaining forecast biases, which might be an indication of a poor fit of the NGR recalibration framework to the data. 
Refinements of the underlying NGR framework are beyond the scope of this paper.

\section{Discussion and conclusions}
\label{sec:summary}

Parameter estimates in forecast recalibration frameworks are subject to uncertainty, particularly when estimated with small training samples.
The effect of parameter uncertainty on the reliability and skill of probability forecasts has received little attention in the climate and meteorology literature.
This study has presented two methods of accounting for parameter uncertainty in recalibrated forecasts.
Analytic results are available for MOS recalibration.
Parameter uncertainty in more complex recalibration frameworks can be estimated by bootstrapping.
The results presented here demonstrate that accounting for parameter uncertainty can improve the reliability and skill of recalibrated forecasts across a range of time scales.

The examples demonstrated here are representative of findings across a number of forecast models, variables and time scales.
In some cases, accounting for parameter uncertainty does not improve forecast reliability and skill. 
For example, accounting for parameter uncertainty did not improve seasonal average European temperature forecasts from the ECMWF System4  \citep{molteni2011system4} at a lead time of 3 months.
The PIT histogram did not change, and the verification scores improved at only 50\% of grid boxes, leaving the average forecast skill unchanged.
However, no cases were identified where accounting for parameter uncertainty leads to forecasts that are less reliable or less skilful on average.

The main effects of accounting for parameter uncertainty are the inflation of the forecast variance, and an increase in the weight of the tails of the forecast distribution.
The wider, heavy tailed forecast distributions improve reliability by generating forecast distributions that are less underdispersive.
This can be seen from the PIT histograms, which are less $\cup$-shaped after accounting for parameter uncertainty.
Overall forecast skill is also improved, but the size of the improvement depends on the score.
The Ignorance score is more sensitive to low-probability events than the CRPS. 
Since the main effect of accounting for parameter uncertainty is to improve the tails of the forecast distributions, the relative improvement in the Ignorance score is larger than that of the CRPS.

The effect of accounting for parameter uncertainty is largest for small training samples.
The amount of training data can be limited by the available observational record, by strong temporal and spatial correlations in the data, or by the computational expense of generating long hindcast experiments.
However, small training samples are often deliberately chosen even in data-rich situations such as weather forecasting.
The use of a rolling training window allows the recalibration to adapt to changes in background conditions, or changes to the forecast model.
Alternative methods might be considered so that all prior data are included in the training sample but with decreasing weight given to older data.
Another possibility would be to use the idea of analogues, and only calibrate using prior data that are similar to conditions observed at the time the forecast is initialized.

Analytic expressions for the forecast uncertainty are possible for some more complex recalibration frameworks.
The predictive bootstrap is easily applicable to almost any recalibration framework.
Bootstrapping can be modified to account for temporal dependence between the training data by block resampling \citep[][Chp. 8]{davison1997bootstrap}.
Alternative methods of estimating the parameter uncertainty include parametric bootstrapping, and asymptotic approximations of the parameter uncertainty.
Bayesian methods lead to similar analytic results in the case of MOS recalibration, and computational Bayesian techniques can be used for more complex frameworks.

This study has demonstrated that accounting for parameter uncertainty in probability forecasts leads to measurable improvements in both reliability and skill.
Other researchers and practitioners are encouraged to investigate and adopt the methods proposed here, and to develop alternative methods for more complex recalibration frameworks.

\section*{Acknowledgments}
The authors would like to thank Adam Scaife and the members of the monthly-to-decadal prediction group at the Met Office Hadley Centre for providing the GloSea5 data.
The authors would also like to thank Nathan Owen, Chris Ferro, David Stephenson, Tom Hamill, and an anonymous reviewer for helpful comments during the preparation of this manuscript.
Stefan Siegert was supported by the European Union Programme FP7/2007-13 under grant agreement 3038378 (SPECS).
Philip Sansom was supported by a grant from the National Oceanic and Atmospheric Administration (NOAA) NA12OAR4310086. 
The views expressed herein are those of the authors and do not necessarily reflect the views of their funding bodies or any of their subagencies.

\section*{Appendices} 

\subsection*{The non-standardized t-distribution}

The pdf of the non-standardized t-distribution with location $\mu$, scale $\sigma$ and degrees-of-freedom $\nu$ is given by
\citep{harrison1999bayesian}
\begin{equation}
  p \left( x; \nu, \mu, \sigma^2 \right) 
    = \frac{\Gamma \left( \frac{\nu + 1}{2} \right)}
           {\Gamma \left( \frac{\nu}{2} \right) \sqrt{\pi \nu \sigma^2}} 
      \left[ 
        1 + \frac{1}{\nu} \frac{\left( x - \mu \right)^2}{\sigma^2}
      \right]^{-\frac{\nu + 1}{2}}.
\end{equation}

\subsection*{Normal mixture distribution}

Let $\varphi(x)$ and $\Phi(x)$ denote the pdf and cdf of the standard Normal distribution respectively.
If the forecast pdf $f(x)$ is a mixture of $K$ Normal distributions with weights $\omega_1, \cdots, \omega_K$ (non-negative and summing to one), means $\mu_1, \cdots, \mu_K$ and variances $\sigma_1^2, \cdots, \sigma_K^2$, then the forecast pdf $f(x)$ is given by
\begin{equation}
  f(x) = \sum_{k=1}^K \omega_k \frac{1}{\sigma_k} 
           \varphi \left( \frac{x - \mu_k}{\sigma_k} \right),
  \label{eqn:normmix-pdf}
\end{equation}
and the forecast cdf $F(x)$ is
\begin{equation}
  F(x) = \sum_{k=1}^K \omega_k \Phi \left( \frac{x - \mu_k}{\sigma_k} \right).
  \label{eqn:normmix-cdf}
\end{equation}

\subsection*{The Ignorance Score}

If the forecast pdf $f(x)$ is Normal with mean $\mu$ and variance $\sigma^2$, then the Ignorance score is given by 
\begin{equation}
  ign(f, y) = \left[ \frac12 \log \left( 2 \pi \sigma^2 \right) + 
                \frac{\left( y - \mu \right)^2}{2 \sigma^2} \right] / \log 2.
\end{equation}
For a mixture of Normals, simply take the negative logarithm of Equation~\ref{eqn:normmix-pdf} after substituting appropriate values for the forecast means and variances and for the observation.

If the forecast pdf $f(x)$ has the form of a non-standardized t-distribution, then the Ignorance score is given by
\begin{multline}
  ign(f, y) = \left[ - \log \Gamma \left( \frac{\nu+1}{2} \right) 
                     + \log \Gamma \left( \frac{\nu}{2} \right) 
                     + \frac12 \log \left( \pi \nu \sigma^2 \right) 
                     \right. \\ \left.
                     + \frac{\nu+1}{2} \log \left( 1 + 
                         \frac{1}{\nu} 
                           \frac{\left( y - \mu \right)^2}{\sigma^2} \right)
              \right] / \log 2.
\end{multline}

\subsection*{The Continuous Ranked Probability Score}

If the forecast cdf $F(x)$ is Normal with mean $\mu$ and variance $\sigma^2$, then the CRPS is given by \citep{gneiting2005calibrated}
\begin{equation}
  crps(F, y) = \sigma \left\{ 
                 \frac{y - \mu}{\sigma} \left[ 
                   2 \Phi \left( \frac{y - \mu}{\sigma} \right) - 1 
                 \right] 
                 + 2 \varphi \left( \frac{y - \mu}{\sigma} \right) 
                 - \frac{1}{\sqrt{\pi}} 
               \right\}.
\end{equation}
\citet{grimit2006the} showed that the CRPS of a mixture of Normal distributions is given by
\begin{multline}
  crps(F, y) = \frac{1}{K} \sum_{k=1}^K \omega_k 
                 A \left( y - \mu_k, \sigma_k^2 \right) \\
             - \frac12 \sum_{k=1}^K \sum_{l=1}^K \omega_k \omega_l 
                 A \left( \mu_k - \mu_l, \sigma_k^2 + \sigma_l^2 \right),
\end{multline}
where 
\begin{equation}
  A \left (\mu, \sigma^2 \right) 
    = 2 \sigma \varphi \left( \frac{\mu}{\sigma} \right) 
    + \mu \left( 2 \Phi \left( \frac{\mu}{\sigma} \right) - 1 \right)
\end{equation}
Analytical results for the CRPS of other distributions are difficult to derive.
For example, no result has been published for the CRPS of the t-distribution, which appears as a predictive distribution in Equation~\ref{eqn:predictive-t}. 
Numerical integration can be used where analytic results do not exist.
The CRPS of forecasts issued as t-distributions was calculated using the function {\tt integrate} as implemented in package {\tt stats} of the {\tt R} statistical computing software \citep[version 3.2.0]{Rmanual}.

\subsection*{Probability Integral Transformations}

If the forecast pdf $f_t(x)$ is Normal with mean $\mu_t$ and standard deviation $\sigma_t$, then the PIT is given by
\begin{equation}
  pit(f_t, y_t)
    = \Phi \left( \frac{ y_t - \mu_t }{\sigma_t} \right).
\end{equation}
The PIT of a weighted sum of Normals is equal to the weighted sum of the PITs of the individual Normals.

If the forecast pdf $f_t(x)$ is issued as a non-standardized t-distribution with location $\mu_t$, scale $\sigma_t$, and $\nu_t$ degrees-of-freedom, then the PIT is given by
\begin{equation}
  pit \left( f_t, y_t \right) = \mathrm{T}_{\nu_t} \left( 
                                    \frac{ y_t - \mu_t }{\sigma_t} 
                                  \right),
\end{equation}
where $T_\nu(x)$ is the cdf of the central t-distribution with $\nu$ degrees-of-freedom.

\bibliographystyle{abbrvnat}
\bibliography{bib}

\end{document}